\documentclass[12pt,english]{article}
\usepackage[T1]{fontenc}
\usepackage[latin1]{inputenc}
\usepackage{a4wide}
\usepackage{float}
\usepackage{graphicx}

\makeatletter

 \newcommand{\lyxaddress}[1]{
   \par {\raggedright #1
   \vspace{1.4em}
   \noindent\par}
 }

\usepackage{setspace}
\usepackage{graphicx}
\usepackage{setspace}
\date{}

\usepackage{babel}
\makeatother
\begin{document}

\title{The Three-Box Paradox Revisited}

\author{Tamar Ravon and Lev Vaidman%
\footnote{vaidman@post.tau.ac.il%
}}

\maketitle

\lyxaddress{School of Physics and Astronomy\\
Raymond and Beverly Sackler Faculty of Exact Sciences\\
Tel-Aviv University, Tel-Aviv 69978, Israel}

\begin{abstract}
The classical Three-Box paradox of Kirkpatrick {[}\emph{J. Phys.} A
\textbf{36} 4891 (2003){]} is compared to the original quantum
Three-Box paradox of Aharonov and Vaidman {[}\emph{J. Phys.} A
\textbf{24} 2315 (1991){]}. It is argued that the quantum Three-Box
experiment is a {}``quantum paradox'' in the sense that it is an
example of a classical task which cannot be accomplished using
classical means, but can be accomplished using quantum devices. It
is shown that Kirkpatrick's card game is analogous to a different
game with a particle in three boxes which does not contain
paradoxical features.
\end{abstract}

\section{Introduction}

The Three-Box paradox is an example presented by Aharonov and
Vaidman \cite{AV1991} of a pre- and post-selected quantum system
that exhibits highly counter-intuitive behavior. It has become the
focus of a long-lived debate, including attempts to demystify and
refute the paradox (see, for example,
\cite{Kastner1999,Griffiths1996,Griffiths1998,Kent1997,GriffithsHartle1998}).
One such attempt in particular, presented by Kirkpatrick
\cite{Kirkpatrick2003a}, involves a {}``classical analogue'' of the
Three-Box paradox in the form of a card-game. In this paper we
analyze the paradoxical features of the Three-Box experiment and
argue that it does not have a classical analogue. We analyze
Kirkpatrick's card game and show that it does not reproduce the
paradoxical features of the original Three-Box experiment.

The paradox in the Three-Box experiment is that at a particular time
we can claim that a particle is in some sense both with certainty
in one box, $A$, and with certainty in another box, $B$. Now, if
a particle is certainly in $A$, then it is certainly not in $B$,
and vice versa. Therefore, if a single particle is both certainly
in $A$ and certainly in $B$ we have a paradox. The difficulty is
not unlike the one presented by a two-slit interference experiment,
where we have to admit that a single particle passes simultaneously
through two separated slits, but it is more acute. In the Three-Box
experiment, the particle is \emph{certain} to be found in $A$ if
searched for in $A$, and \emph{certain} to be found in $B$ if searched
for in $B$ instead.

Today we do not know of any real {}``paradoxes'' in physics. A true
physical paradox would be a prediction of a current physical theory
that contradicts experimental results. Such a paradox would
necessitate a modification of the current theory, i.e. progress in
physics. What we mean by {}``quantum paradox'' is a phenomenon that
\emph{classical} physics cannot explain. The Three-Box paradox
describes a particular task that cannot be accomplished using
classical means.

An example of this type of quantum paradox is the Greenberger-Horne-Zeilinger
\cite{GHZ1989} (GHZ) setup modified by Mermin \cite{Mermin1990},
in which a team equipped with quantum technology can win with certainty
a game with separated players that can be won by a team employing
classical (relativistic) physics only with probability 0.75 \cite{Vaidman1999}.
The GHZ game is one of numerous games \cite{Vaidman2002,KwiatHardy2000}
based on quantum entanglement, a phenomenon that enables correlations
between remote observers that are stronger than those possible classically
\cite{Bell1965}. The Three-Box experiment is one of the only quantum
paradoxes which do not employ entanglement between remote locations,
so it is a genuinely new paradox%
\footnote{One can also consider Interaction-Free Measurement \cite{EV1993IFM}
as a quantum paradox which is not based on classically impossible
correlations. There, a gedanken device is considered which explodes
whenever a particle {}``touches'' it. Quantum strategy allows to
detect the device without it exploding.%
}. The existence of a classical system which can perform the Three-Box
task would remove the paradox, and thus it is important to analyze
the validity of Kirkpatrick's argument.

In the next section we consider some unusual properties of a pre-
and post-selected spin-$\frac{1}{2}$ particle, in order to clarify
the main paradoxical feature of the quantum Three-Box experiment described
later in Section 3. Section 4 is devoted to a presentation of Kirkpatrick's
game as given in his paper. In section 5 we strip Kirkpatrick's game
of those elements which are not relevant to the analogy with the Three-Box
paradox and describe other simple classical games with similar properties.
This allows us to show that Kirkpatrick's game does not present a
classical analogue to the quantum Three-Box paradox. Section 6 concludes
the paper with a broader view on related issues.

\section{Paradoxical Features of the Pre- and Post-selected Spin-$\frac{1}{2}$
Particle}

In quantum theory, unlike classical theory, an initial (or final)
state alone does not provide all the information about a system during
the time interval between two measurements. For pre- and post-selected
quantum systems, a complete description is given by a two-state vector
\cite{AV1990} that takes into account both initial and final states.
This is the basis for all the examples in which we can make apparently
contradictory statements about a quantum system at the time between
its pre- and post-selection.

For example, consider a spin-$\frac{1}{2}$ particle pre-selected
in the state $\left|\uparrow_{x}\right\rangle $ and post-selected
in the state $\left\langle \uparrow_{z}\right|$. An intermediate
measurement of $\sigma_{x}$ or of $\sigma_{z}$ is in either case
certain to yield $+1$. The \emph{}spin \emph{}of the particle is
\emph{certain} to be {}``up'' in the $x$ direction if a projection
measurement onto $\left|\uparrow_{x}\right\rangle $ is performed,
and \emph{certain} to be {}``up'' in the $z$ direction if a projection
measurement onto $\left|\uparrow_{z}\right\rangle $ is performed.

A naive classical analogy of a spin {}``up'' in the $x$ direction
is a classical pointer, an arrow, pointing in the $x$ direction.
An arrow pointing in the $x$ direction certainly does not point in
the $z$ direction, so apparently we have obtained a paradox of the
kind associated with pre- and post-selection. However, it is not so,
since there is no classical task involved. Spin measurement is genuinely
quantum, and within the framework of quantum theory there is no paradox.
For, when the spin is in the state $\left|\uparrow_{x}\right\rangle $,
we cannot claim that it is not pointing in the $z$ direction; even
without post-selection there is a 50\% chance of finding it in the
state $\left|\uparrow_{z}\right\rangle $.

An example of a situation that is paradoxical within the framework
of quantum theory is a pre- and post-selected spin-$\frac{1}{2}$
particle that is certain to be {}``up'' in the $x$ direction, certain
to be {}``up'' in the $y$ direction, \emph{and} certain to be {}``up''
in the $z$ direction \cite{VAA1987}. Here we discuss another paradoxical
experiment; it is based on the mathematical structure of the Three-Box
paradox. Would you be surprised by the following? Suppose that we
give you a box, and arrange a \textit{single} pre- and post-selected
particle so that:

\begin{quote}
If the particle is searched for in a box with its spin {}``up''
in the $z$ direction, it is to be found with certainty, while if
the same particle is searched for in the same box with its spin {}``down''
in the $z$ direction, it is also to be found with certainty.
\end{quote}
These two properties are contradictory not only in the naive classical
analogue (of an arrow pointing in two opposite directions) but also
in standard quantum theory. The usual {}``eigenvalue - eigenstate''
link suggests that a particle which is to be found with certainty,
if searched for, in a box with its spin {}``up'' in the $z$ direction
has the quantum state $\left|\uparrow_{z}\right\rangle $, and thus
cannot be found in the box with its spin {}``down'' in the $z$
direction.

It is possible for a pre- and post-selected system to yield definite
outcomes for intermediate measurements that are not ensured by either
the pre-selection or the post-selection alone. Still, a situation
in which contradictory outcomes are obtained with certainty, if measured,
is surprising. Let us show how this is achieved.

The system consists of a spin-$\frac{1}{2}$ particle and two boxes,
$A$ and $B$. The particle is pre-selected in the state\begin{eqnarray}
\left|\psi\right\rangle  & = & \frac{1}{\sqrt{3}}\left(\left|A,\uparrow_{z}\right\rangle +\left|A,\downarrow_{z}\right\rangle +\left|B,\uparrow_{z}\right\rangle \right)\,\,\,,\label{eq:spinpre}\end{eqnarray}
 and post-selected in the state \begin{eqnarray}
\left\langle \phi\right| & = & \frac{1}{\sqrt{3}}\left(\left\langle A,\uparrow_{z}\right|+\left\langle A,\downarrow_{z}\right|-\left\langle B,\uparrow_{z}\right|\right)\,\,\,,\label{eq:spinpost}\end{eqnarray}
 where $\left|A,\uparrow_{z}\right\rangle $ represents the particle
in box $A$ with spin {}``up'' in the $z$ direction. We give you
box $A$ and offer you to either look for a particle with spin {}``up''
or to look for a particle with spin {}``down'' inside. To do this,
you have to perform an ideal projection measurement. Therefore, if
you look for the particle with spin {}``up'' and find it, the particle's
quantum state after your measurement becomes $\left|A,\uparrow_{z}\right\rangle $,
but if you do not find it, its state becomes $\frac{1}{\sqrt{2}}(\left|A,\downarrow_{z}\right\rangle +\left|B,\uparrow_{z}\right\rangle )$.
The latter is orthogonal to the post-selected state, however, so the
post-selection cannot succeed if the particle is not found. And because
the pre- and post-selections are symmetric with respect to the states
$\left|A,\uparrow_{z}\right\rangle $ and $\left|A,\downarrow_{z}\right\rangle $,
the particle is also to be found, if searched for, with its spin {}``down''
in the same box.

Although this spin example is more counter-intuitive than the previous
one, it too does not fall into the category of {}``quantum paradoxes''
as defined above, since there is no classical task involved. In both
examples we discuss the results of spin measurements which are inherently
quantum. The measurements in the second example are particularly difficult.
In the first example, a suitably oriented Stern-Gerlach apparatus
can be used, but in the second example this is no longer appropriate
- a Stern-Gerlach measurement in box $A$ would distinguish between
all three states $\left|A,\uparrow_{z}\right\rangle $, $\left|A,\downarrow_{z}\right\rangle $,
and $\left|B,\uparrow_{z}\right\rangle $, and therefore would not
be a projection measurement.

\section{The Three-Box Paradox}

The original Three-Box experiment \cite{AV1991} involves no spin,
only a particle and three separate boxes. The particle is pre-selected
in the state\begin{eqnarray}
\left|\psi\right\rangle  & = & \frac{1}{\sqrt{3}}\left(\left|A\right\rangle +\left|B\right\rangle +\left|C\right\rangle \right)\,\,\,,\label{eq:3boxpre}\end{eqnarray}
 and post-selected in the state \begin{eqnarray}
\left\langle \phi\right| & = & \frac{1}{\sqrt{3}}\left(\left\langle A\right|+\left\langle B\right|-\left\langle C\right|\right)\,\,\,,\label{eq:3boxpost}\end{eqnarray}
 where the mutually orthogonal states $\left|A\right\rangle $, $\left|B\right\rangle $,
and $\left|C\right\rangle $ denote the particle being in box $A$,
$B$, and $C$, respectively.

In between pre- and post-selection, an observation takes place either
of box $A$ or of box $B$. A successful observation of box $A$ corresponds
to the particle being found in the box and is represented by the projection
$\mathrm{P}_{A}=\left|A\right\rangle \left\langle A\right|$. An unsuccessful
observation of box $A$ corresponds to the particle \emph{not} being
found and is represented by $\mathbf{1}-\mathrm{P}_{A}=\left|B\right\rangle \left\langle B\right|+\left|C\right\rangle \left\langle C\right|$.
Similar definitions hold for an observation of box $B$. As before,
each of the two intermediate observations is certain to succeed. Indeed,
not finding the particle in $A$ leads to the quantum state $\frac{1}{\sqrt{2}}\left(\left|B\right\rangle +\left|C\right\rangle \right)$,
and not finding it in $B$ leads to the state $\frac{1}{\sqrt{2}}\left(\left|A\right\rangle +\left|C\right\rangle \right)$,
both of which are orthogonal to the post-selected state $\left\langle \phi\right|$.

The Three-Box example can be presented as a classical task which cannot
be achieved using classical means but which can be achieved using
quantum preparation and verification measurements. Consider the following
game. Alice, equipped with quantum devices, prepares the particle
in the system of three boxes and passes the first two boxes on to
Bob. Bob, who is unaware of Alice's quantum machinery, is told to
look in one of the boxes, and that he wins if he does \emph{not} find
anything. However, Alice gets to decide according to her post-selection
measurement whether the game {}``counts'' or not. Now, Bob has no
\emph{a priori} reason not to agree to play; his chances of winning
appear to be 1/2. It is obviously so before Alice performs the post-selection
measurement, and since Bob is careful not to leave any mark disclosing
the results of his measurement, Alice apparently cannot gain anything
from the post-selection. Bob will find, however, that Alice somehow
manages to discard all runs of the game in which he does not find
the particle, and so Bob will always lose.

\section{Kirkpatrick's Game}

Kirkpatrick \cite{Kirkpatrick2003a} claims that {}``...the Three-Box
example is neither quantal nor a paradox...'', and supposedly demonstrates
this by means of its reproduction in an elaborate playing card game.

Kirkpatrick's classical system is constructed using ordinary playing
cards. Each playing card has two marks, a {}``face'' (e.g. Jack,
Queen, King) and a {}``suit'' (e.g. Spades, Diamonds, Hearts), and
these are treated as system variables: $Face$ (with values $J$,
$Q$, $K$) and $Suit$ (with values $S$, $D$, $H$). The deck of
cards is divided into two parts called $These$ and $Others$, and
the system consists also of a memory $\mathcal{M}$. Kirkpatrick defines
the following procedures, in which a system variable is generically
referred to as $P$ and its values as $\{ p_{j}\}$.

\paragraph{Preparation.}

The system is {}``prepared'' in a particular {}``state'' $P=p_{j}$
(e.g. $Face=Q$) by (1) placing all the cards with $P=p_{j}$ (i.e.
all the Queens) in $These$ and the remainder of the deck in $Others$,
and (2) setting the memory to the variable name $\mathcal{M}\equiv P$
(e.g $\mathcal{M}\equiv Face$).

\paragraph{Observation.}

The variable $P$ is {}``observed'' in the following manner. If
$\mathcal{M}=P$, (1) select a card at random from $These$ and (2)
report the value $p_{j}$ of the variable $P$ of this card (i.e.,
its {}``face'' or {}``suit''). If $\mathcal{M}\neq P$, then (1)
select a card at random from $Others$, (2) report the value $p_{j}$
of the variable $P$ of this card, and (3) prepare the system in the
state $P=p_{j}$ according to the procedure defined above.

\paragraph{Partial Observation.}

In order to reproduce the Three-Box experiment, it should be possible
to perform a partial (as opposed to complete) observation, in which
one observes only whether or not the variable $P$ has a particular
value $p_{j}$. In a partial measurement of $P$ one selects a card
from the appropriate part of the deck (as determined by the content
of $\mathcal{M}$) and reports the value $p_{j}$ if the card's value
of $P$ is indeed $p_{j}$, and $\widetilde{p_{j}}$ if the card's
value of $P$ is \emph{not} $p_{j}$. If $\mathcal{M}\neq P$, one
prepares the system according to the outcome. To prepare the system
in the state $P=\widetilde{p_{j}}$, place all the cards with $P=\widetilde{p_{j}}$
in $These$ and all other cards (i.e., those with $P=p_{j}$) in $Others$.
The variable $\mathcal{M}$ is, as before, set to $\mathcal{M}\equiv P$.\\

The role of the variable $\mathcal{M}$ is apparently to distinguish
between {}``repeated'' and {}``new'' observations. A repeated
measurement is certain to yield the same outcome as obtained previously,
whereas a {}``new'' measurement (which by definition must be a measurement
of the other variable) causes the system's state to be reset.

The need to define partial observations has to do with the Three-Box
analogue. In the Three-Box experiment, projection measurements are
involved that measure only whether or not the particle is in a particular
box. Such measurements are considered to be {}``partial'', whereas
a {}``complete'' measurement would consist of, e.g., looking in
all three boxes.

The Three-Box analogue is supposedly obtained by choosing the following
six cards for the deck: a Jack of Spades ($JS$), a Jack of Diamonds
($JD$), a Queen of Spades ($QS$), a Queen of Diamonds ($QD$), and
two Kings of Hearts ($(2)KH$). (The reason for two Kings of Hearts
is to have each value of each variable appear the same number of times
in the deck). The system is prepared in the state $Face=Q$. One of
two partial observations of the variable $Suit$ then takes place:
either of whether or not $Suit=S$ (corresponding to an observation
of box $A$) or of whether or not $Suit=D$ (corresponding to an observation
of box $B$). Finally, a post-selection measurement of $Face$ is
performed resulting in the final state $Face=K$.

The initial state $Face=Q$ ensures that all outcomes are possible
in the following observation of $Suit$ (since the $JS$, the $JD$,
and the $(2)KH$ are in $Others$). It is the post-selected, final
state that ensures that if the intervening observation was of whether
$Suit=S$, then the $JS$ must have been selected, and if it was of
whether $Suit=D$, then the $JD$ must have been selected. This is
because in the final measurement of $Face$ a $KH$ must be selected
from $Others$, and whereas in the states $Suit=S$ and $Suit=D$
the $(2)KH$ remain in $Others$, in the states $Suit=\widetilde{S}$
and $Suit=\widetilde{D}$ they are moved to $These$. Therefore, given
the post-selection, the $JS$ and the $JD$ are each certain to be
selected, and the system's intermediate state is certain to be $Suit=S$
and certain to be $Suit=D$, depending on what is measured. This,
claims Kirkpatrick, is analogous to the particle in the Three-Box
experiment being found with certainty either in box $A$ or in box
$B$. %
\begin{figure}[H]
\includegraphics[%
  bb=0bp 280bp 594bp 820bp,
  clip,
  scale=0.67]{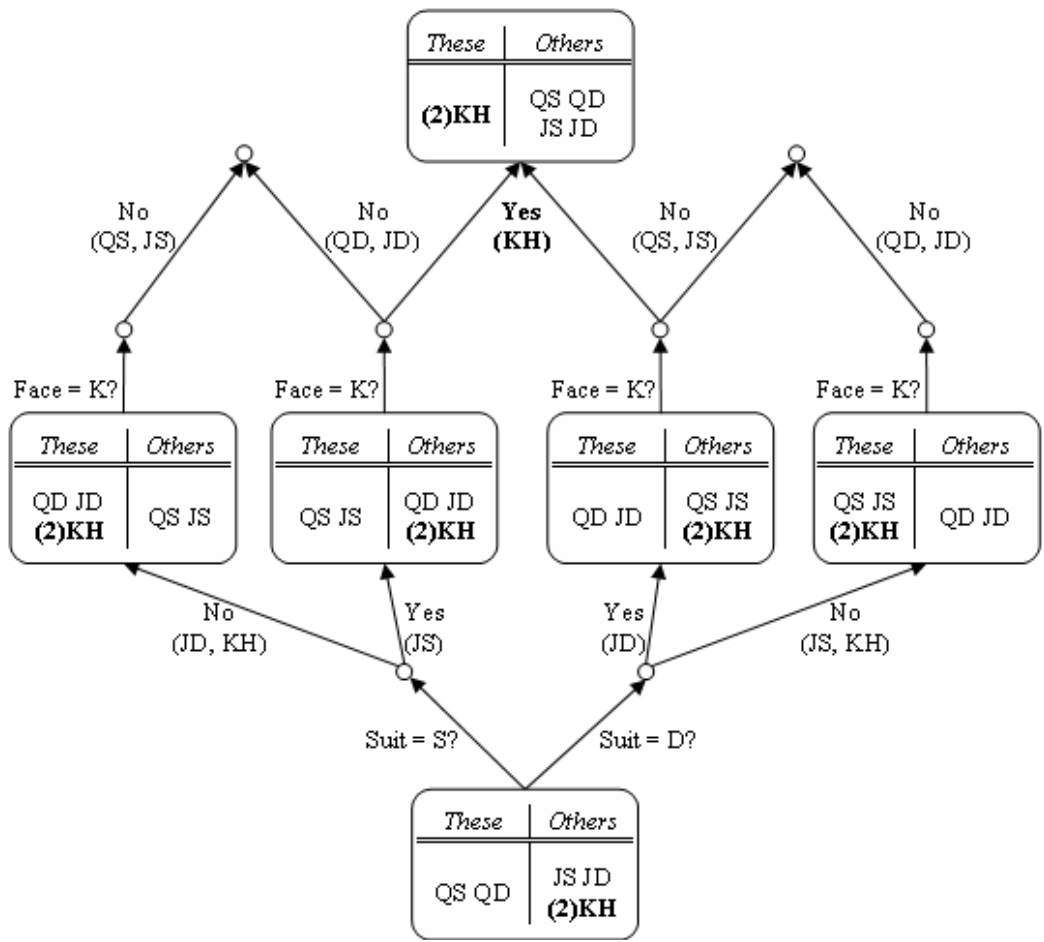}

\caption{Kirkpatrick's card-game. The initial \emph{Face=Q} state ensures
that all outcomes are possible in the subsequent measurement of \emph{Suit}.
The final \emph{Face=K} state requires that a \emph{KH} be selected
from \emph{Others} in the final measurement of \emph{Face}. This,
in turn, means that a \emph{JS} must be selected in a measurement
of whether \emph{Suit=S}, and a \emph{JD} must be selected in a measurement
of whether \emph{Suit=D}.}
\end{figure}

\section{Simplification of Kirkpatrick's Game and Similar Proposals}

The large part of Kirkpatrick's game is extraneous to the Three-Box
analogy. It is intended to mimic quantum mechanical phenomena in a
more general sense. In another work \cite{Kirkpatrick2003b}, Kirkpatrick
uses a modified version of the game presented here, claiming that
it illustrates the ordinary nature of much of quantum probability,
including incompatibility of observables, interference, etc. We are
not pursuaded by his arguments, but the discussion of these issues
goes beyond the scope of this paper. We therefore consider Kirkpatrick's
game without the complications of defining (and using the terminology
of) system states and variables.

Consider a smaller deck of cards consisting of a Jack of Spades ($JS$),
a Jack of Diamonds ($JD$), and a King of Hearts ($KH$), still distributed
between two groups $These$ and $Others$. Suppose that initially
all three cards are in $Others$. A card is then selected at random
from $Others$, and as before a {}``partial observation'' is performed
either of whether or not it is a spade (i.e., the $JS$), or of whether
or not it is a diamond (the $JD$). In either case, if the outcome
is positive, the selected card is placed in $These$ while the others
remain in $Others$. Otherwise, the selected card is returned to $Others$
while the other cards are moved to $These$. Finally, a card is again
selected at random from $Others$, and our post-selection requirement
is that this final card is the $KH$. This requirement ensures that
the previously selected card must have been the $JS$ if spade was
searched for, and must have been the $JD$ if diamond was searched
for instead.

\begin{figure}[H]
\includegraphics[%
  bb=0bp 360bp 594bp 825bp,
  scale=0.67]{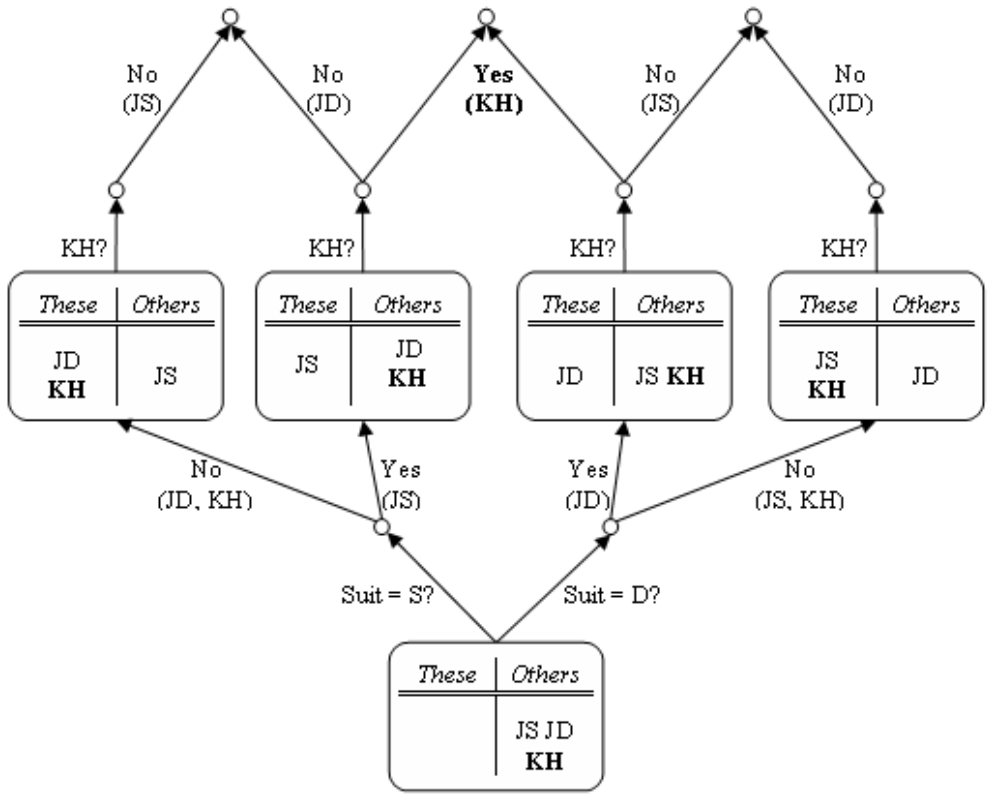}

\caption{Simplification of Kirkpatrick's game. Initially three cards \emph{JS},
\emph{JD}, and \emph{KH} are in \emph{Others}, allowing for any of
the relevant values of \emph{Suit} to be selected. The final, post-selection
requirement is that a \emph{KH} be selected from \emph{Others}. Therefore
a \emph{JS} must be selected if it is searched for, and that a \emph{JD}
must be selected if it is searched for instead.}
\end{figure}

Leifer and Spekkens \cite{LeiferSpekkens2005} have suggested yet
a simpler game to demonstrate this classical phenomenon. Their system
consists of a ball and a box. The box can be split into two half-boxes,
either length-wise or width-wise, by placing a double partition inside
and separating the resulting halves. The two half-boxes can be reassembled
into a single box by joining them together and removing the partition.
The halves of one division are denoted {}``front'' and {}``back'',
and of the other, {}``right'' and {}``left''. One measures whether
the ball is in the front half of the box by partitioning the box into
front and back halves and shaking the front half to hear whether the
ball is inside. If the ball is found in this case, its left/right
position is randomized, but if it is not found then its left/right
position is undisturbed. Similar definitions apply for measurements
of whether the ball is in the back, on the right, or on the left.
Now suppose that this system is pre-selected so that the ball is in
the front half, and post-selected so that the ball is in the back
half, where the intervening measurement consists of looking for the
ball either on the right or on the left. Then, the ball is necessarily
found in the intervening measurement, since otherwise there is no
disturbance of the initial front state and no way for the ball to
be transferred to the back half. Thus, the ball is certain to be found
on the right if that is where it is looked for, and certain to be
found on the left if that is where it is looked for instead. %
\begin{figure}[H]
\includegraphics[%
  bb=0bp 380bp 594bp 825bp,
  scale=0.67]{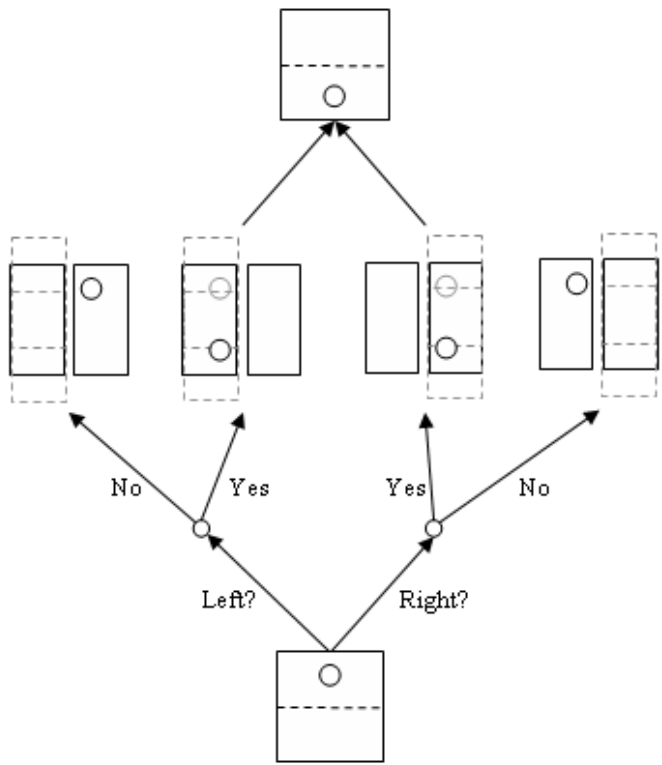}

\caption{Leifer and Spekkens' game. The ball is pre-selected in the front
half and post-selected in the back half. Its movement from the front
to the back ensures that the ball is found in the intervening measurement
either of whether it is on the left or of whether it is on the right.}
\end{figure}

What these games have in common is that they use classical measurement
disturbance to effectively encode the outcome of the measurement into
the system. In both cases, parts of the system are moved around in
such a way that the post-selection becomes impossible whenever the
measurement is unsuccessful. Obviously there is nothing paradoxical
about such mechanisms, but neither are they equivalent to the Three-Box
experiment. In the Three-Box experiment, the intermediate observation
consists of just that - observation. A classical system, on the other
hand, is not disturbed by observation, therefore the measurement must
involve additional actions.

We can suggest another Three-Box experiment, or game, which utilizes
this kind of classical measurement disturbance and is entirely non-paradoxical.
Suppose we put a classical ball in one of three boxes, and have another
observer look inside either the first box or the second. So long as
the box is only observed, there is no post-selection measurement we
can perform that will ensure the ball has been found. Therefore, let
us have the observer place the ball in the third box whenever he does
not find it in the observed box. This makes it is easy to ensure the
ball is found by post-selecting that finally the ball is not in the
third box. It is this Three-Box game, rather than the original one,
to which Kirkpatrick's game is analogous. %
\begin{figure}[H]
\includegraphics[%
  bb=0bp 360bp 594bp 825bp,
  scale=0.67]{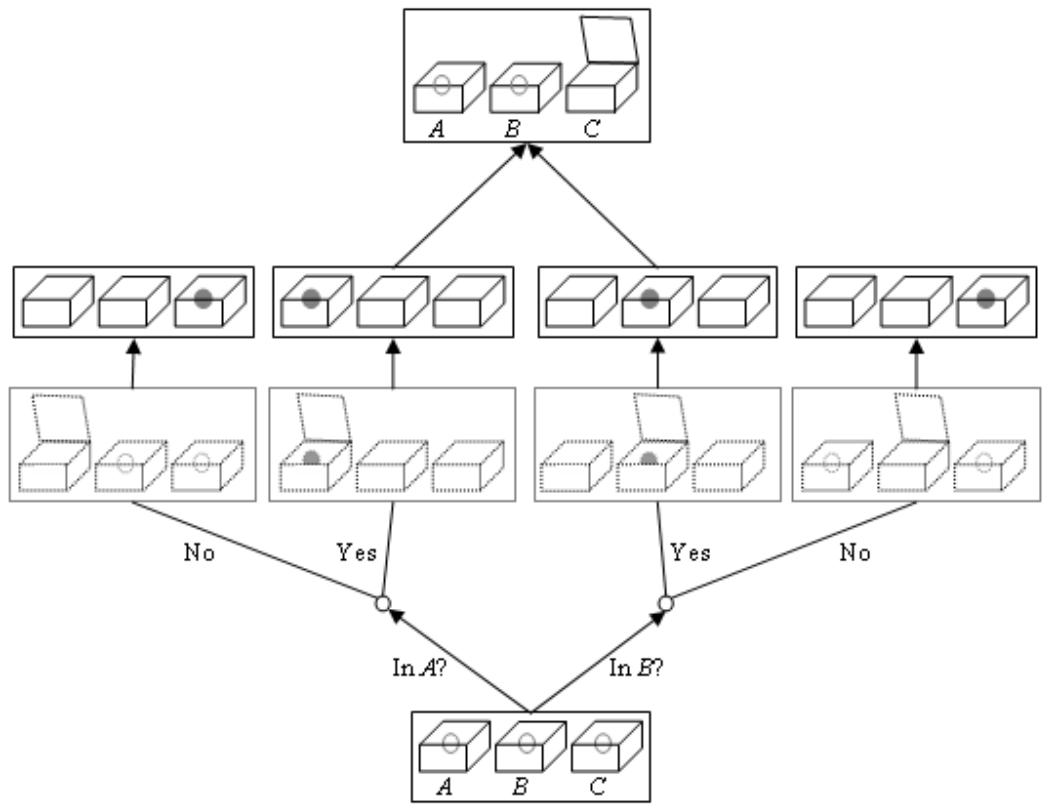}

\caption{The Three-Box game to which Kirkpatrick's game is analogous. Initially
the ball is in one of the three boxes. If the observer does not find
the ball in one of the first two, he moves the ball to the third box.
The post-selection is that the ball is not in the third box.}
\end{figure}

In all these games the {}``observation'' includes leaving a mark
that is later {}``read'' in the post-selection process. The observation
in the original Three-Box experiment, which consists of opening a
box, does not leave such a mark in the framework of classical physics.
Note that, contrary to the supposedly analogous games, a successful
post-selection in the Three-Box experiment is possible even with no
intermediate measurement (i.e., if none of the boxes are opened).

\section{Quantum Paradox and Beyond}

The quantum Three-Box experiment, beyond its paradoxality, is very
effective in demonstrating concepts that are related specifically
to pre- and post-selected quantum systems. While we agree (see \cite{Vaidman1999Reply})
with Kastner \cite{Kastner1999} that in the Three-Box experiment
there is no ontological property, or any kind of hidden variable,
corresponding to the particle {}``being in box $A$'', we believe
that consistent and useful operational definitions can be obtained
for the {}``elements of reality'' \cite{Vaidman1993ER} and {}``weak-measurement
elements of reality'' \cite{Vaidman1996WMER} associated with the
particle {}``being in box $A$''. The former corresponds to {}``if
observed in $A$, is to be found there with certainty'', and the
latter to the \textit{weak value} associated with a \emph{weak measurement}
\cite{AV1990} of the particle in $A$.

In the case that we observe the particle in $A$ weakly (i.e., with
negligible disturbance), these concepts become particularly useful,
for it is then possible to weakly observe the particle in $B$ as
well, at the same time. As it turns out, the {}``elements of reality''
and {}``weak-measurement elements of reality'' are the same as if
there were indeed two particles, one in $A$ and another in $B$ \cite{RLS2004Exp}.

As we have shown, Kirkpatrick's system does not reproduce the Three-Box
paradox because it contains no contradiction. In \cite{Kastner1999},
Kastner claims that the original Three-Box experiment does not contain
a contradiction either. Her argument is that because only one of the
boxes is actually observed, the properties of {}``being in box $A$''
and of {}``being in box $B$'' cannot be attributed simultaneously
to the same single particle. Kastner argues that once the particle
is found, one cannot consider what would have happened had the other
box been observed instead. This implies that when the particle is
found it actually becomes {}``in the box'' as a result of the observation,
i.e., the measurement disturbance. Thus, although Kastner does not
dispute the quantum mechanical nature of the Three-Box experiment,
Kirkpatrick's game seems to provide a good demonstration of her argument.

Kastner and Kirkpatrick apparently have the following idea in
common: that if disturbance is understood to be inherent to
measurement, then the difficulty with regard to the Three-Box
experiment is removed. This view is shared by Leifer and Spekkens,
who show \cite{LeiferSpekkens2005} that any attempt to create a
classical analogue of the Three-Box paradox requires measurement
disturbance. But it is precisely because the classical observation
in the Three-Box experiment is \emph{non-disturbing} that the
experiment cannot be explained by classical physics or accomplished
using classical means. This is what causes the Three-Box experiment
to be a {}``quantum paradox''.

This work has been supported by the European Commission under the
Integrated Project Qubit Applications (QAP) funded by the IST directorate
as Contract Number 015848.


\begin{thebibliography}{10}
\bibitem{AV1991}Aharonov Y and Vaidman L 1991 {}``Complete Description of a quantum
system at a given time'' \emph{J. Phys.} A \textbf{24} 2315-28.
\bibitem{Kastner1999}Kastner R E 1999 {}``The Three-Box {}``Paradox'' and other reasons
to reject the counterfactual usage of the ABL rule'' \emph{Found.
Phys.} \textbf{29}(6) 851-63.
\bibitem{Griffiths1996}Griffiths R B 1996 {}``Consistent histories and quantum reasoning''
\emph{Phys. Rev.} A \textbf{54} 2759-74.
\bibitem{Griffiths1998}Griffiths R B 1998 {}``Choice of consistent family, and quantum incompatibility''
\emph{Phys. Rev.} A \textbf{57} 1604-18.
\bibitem{Kent1997}Kent A 1997 {}``Consistent sets yield contrary inferences in quantum
theory,'' \emph{Phys. Rev. Lett.} \textbf{78} 2874-7.
\bibitem{GriffithsHartle1998}Griffiths R B and Hartle J B 1998 {}``Comment on {}``Consistent
sets yield contrary inferences in quantum theory'''' \emph{Phys.
Rev. Lett.} \textbf{81} 1981.
\bibitem{Kirkpatrick2003a}Kirkpatrick K A 2003 {}``Classical Three-Box {}``paradox'''' \emph{J.
Phys.} A \textbf{36}(17) 4891-900.
\bibitem{GHZ1989}Greenberger D M, Horne M A, and Zeilinger A 1989 in \emph{Bell's Theorem,
Quantum Theory, and Conceptions of the Universe} ed M Kafatos (Dordrecht:
Kluwer Academic) p 69.
\bibitem{Mermin1990}Mermin N D 1990 {}``Quantum mysteries revisited'' \emph{Am. J. Phys.}
\textbf{58} 731.
\bibitem{Vaidman1999}Vaidman L 1999 {}``Variations on the theme of the Greenberger-Horne-Zeilinger
proof'' \emph{Found. Phys.} \textbf{29} 615-630.
\bibitem{Vaidman2002}Vaidman L 2002 {}``An impossible necklace'' in \emph{Quantum {[}Un{]}speakables:
From Bell's Theorem to Quantum Information} ed R Bertlmann and A Zeilinger
(Berlin: Springer) pp 221-4.
\bibitem{KwiatHardy2000}Kwiat P G and Hardy L 2000 {}``The mystery of the quantum cakes''
\emph{Am. J. Phys.} \textbf{68} 33.
\bibitem{Bell1965}Bell J S 1965 {}``On the Einstein-Podolsky-Rosen paradox'' \emph{Physics}
\textbf{1} 195.
\bibitem{EV1993IFM}Elitzur A C and Vaidman L 1993 {}``Quantum mechanical interaction-free
measurements'' \emph{Found. Phys.} \textbf{23}(7) 987-97.
\bibitem{AV1990}Aharonov Y and Vaidman L 1990 {}``Properties of a quantum system
during the time interval between two measurements'' \emph{Phys. Rev.}
A \textbf{41} 11-20.
\bibitem{VAA1987}Vaidman L, Aharonov Y, and Albert D Z (1987) {}``How to ascertain
the values of $\sigma_{x}$, $\sigma_{y}$, and $\sigma_{z}$ of a
spin-$1/2$ particle'' \emph{Phys. Rev. Lett.} \textbf{58} 1385-87.
\bibitem{Kirkpatrick2003b}Kirkpatrick K A 2003 {}``''Quantal'' behavior in classical probability''
\emph{Found. Phys. Lett.} \textbf{16} 199-224.
\bibitem{LeiferSpekkens2005}Leifer M S and Spekkens R W 2005 {}``Logical pre- and post-selection
paradoxes, measurement-disturbance and contextuality'' \textit{Int.
J. Theor. Phys.} \textbf{44} 1977-87.
\bibitem{Vaidman1999Reply}Vaidman L 1999 {}``The meaning of elements of reality and quantum
counterfactuals: Reply to Kastner'' \emph{Found. Phys.} \textbf{29}(6)
865-76.
\bibitem{Vaidman1993ER}Vaidman L 1993 {}``Lorentz-invariant {}``elements of reality''
and the joint measurability of commuting observables'' \emph{Phys.
Rev. Lett.} \textbf{70} 3369-72.
\bibitem{Vaidman1996WMER}Vaidman L 1996 {}``Weak-measurement elements of reality'' \emph{Found.
Phys.} \textbf{26}(7) 895-906.
\bibitem{RLS2004Exp}Resch K J, Lundeen J S, and Steinberg A M 2004 {}``Experimental realization
of the quantum box problem'' \emph{Phys. Lett.} A \textbf{324}(2-3)
125-31.\end{thebibliography}
\end{document}